\documentclass[]{spie}  %>>> use for US letter paper
\usepackage{amsmath,amsfonts,amssymb}
\usepackage{graphicx}
\usepackage{pifont}
\usepackage{epstopdf}
\usepackage[colorlinks=true, allcolors=blue]{hyperref}

\title{Dual Network Architecture for Few-view CT -- Trained on ImageNet Data and Transferred for Medical Imaging}

\author[a]{Huidong Xie}
\author[a]{Hongming Shan}
\author[a]{Wenxiang Cong}
\author[b]{Xiaohua Zhang}
\author[b]{Shaohua Liu}
\author[b]{Ruola Ning}
\author[a]{Ge Wang}
\affil[a]{Biomedical Imaging Center, Department of Biomedical Engineering, Center for Biotechnology \& Interdisciplinary Studies, Rensselaer Polytechnic Institute, 110 Eighth Street, Troy NY 12180, USA}
\affil[b]{Koning Corporation, West Henrietta NY 14586, USA}

\authorinfo{Corresponding author: Ge Wang. E-mail: wangg6@rpi.edu}

\begin{document} 
\maketitle
\begin{abstract}
X-ray computed tomography (CT) reconstructs cross-sectional images from projection data. However, ionizing X-ray radiation associated with CT scanning may induce cancer and genetic damage. Therefore, the reduction of radiation dose has attracted major attention. Few-view CT image reconstruction is an important topic to reduce the radiation dose. Recently, data-driven algorithms have shown great potential to solve the few-view CT problem. In this paper, we develop a dual network architecture (DNA) for reconstructing images directly from sinograms. In the proposed DNA method, a point-wise fully-connected layer learns the backprojection process requesting significantly less memory than the prior arts do. Proposed method uses $O(C\times N\times N_c)$ parameters where $N$ and $N_c$ denote the dimension of reconstructed images and number of projections respectively. $C$ is an adjustable parameter that can be set as low as 1. Our experimental results demonstrate that DNA produces a competitive performance over the other state-of-the-art methods. Interestingly, natural images can be used to pre-train DNA to avoid overfitting when the amount of real patient images is limited.
\end{abstract}


\section{INTRODUCTION}
\label{sec:intro}  % \label{} allows reference to this section

Few-view CT is often mentioned in the context of tomographic image reconstruction. Because of the requirement imposed by the Nyquist sampling theorem, reconstructing high-quality CT images from under-sampled data was considered impossible. When sufficient projection data are acquired, analytical methods such as filtered backprojection (FBP)\cite{wang_approximate_2007} are widely used for clinical CT image reconstruction. In few-view CT circumstance, severe streak artifacts are introduced in these analytically reconstructed images due to the incompleteness of projection data. To overcome this issue, various iterative techniques were proposed in the past decades, which can incorporate prior knowledge in the image reconstruction. Well-known methods include algebraic reconstruction technique (ART) \cite{gordon_algebraic_1970}, simultaneous algebraic reconstruction technique (SART) \cite{andersen_simultaneous_1984}, expectation maximization (EM) \cite{dempster_maximum_1977}, etc. Nevertheless, these iterative methods are time-consuming and still not able to produce satisfying results in many cases. Recently, the development of the graphics processing unit (GPU) technology and the availability of big data allow researchers to train deep neural networks in an acceptable amount of time. Therefore, deep learning has become a new frontier for CT reconstruction research \cite{wang_perspective_2016, wang_guest_2015,wang_image_2018}.

In the literature, only a few deep learning methods were proposed for reconstructing images directly from raw data. Zhu et al. \cite{zhu_image_2018} use fully-connected layers to learn the mapping from raw k-space data to a corresponding reconstructed MRI image. There is no doubt that fully-connected layers can be used to learn the mapping from the sinogram domain to the image domain. However, importing the whole sinograms into the network requires a significant amount of memory and posts a major challenge to train the network for a full-size CT image/volume on a single consumer-level GPU such as an NVIDIA Titan Xp. A recently proposed method, iCT-Net \cite{li_learning_2019} reduces the computational complexity from $O(N^4)$ in \cite{zhu_image_2018} to $O(N^2\times N_d)$, where $N$ and $N_d$ denote the size of medical images and the number of detectors respectively. But one consumer-level GPU is still unable to handle the iCT-Net.

In this study, we propose a dual network architecture (DNA) for CT image reconstruction, which reduces the required parameters from $O(N^2\times N_d)$ of iCT-Net to $O(C\times N\times N_c)$, where $C$ is an adjustable hyper-parameter much less than $N$, which can be even set as low as 1. Theoretically, the larger the $C$, the better the performance. The proposed network is trainable on one consumer-level GPU such as NVIDIA Titan Xp or NVIDIA 1080 Ti. The proposed DNA is inspired by the FBP formulation to learn a refined filtration backprojection process for reconstructing images directly from sinograms. For X-ray CT, every single point in the sinogram domain only relates to pixels/voxels on an X-ray path through a field of view. With this intuition, the reconstruction process of DNA is learned in a point-wise manner, which is the key ingredient in DNA to alleviate the memory burden. Also, insufficient training dataset is another major issue in deep imaging. Inspired by \cite{zhu_image_2018}, we first pre-train the network using natural images from the ImageNet \cite{noauthor_ImageNet_nodate} and then fine-tune the model using real patient data. To our best knowledge, this is the first work using ImageNet images to pre-train a medical CT image reconstruction network. In the next section, we present a detailed explanation for our proposed DNA network. In the third section, we describe the experimental design, training data and reconstruction results. Finally, in the last section, we discuss relevant issues and conclude the paper.

\section{METHODS}
This section presents the proposed dual network architecture and the objective functions.
\subsection{Dual network architecture (DNA)}\label{sec:2.1}

DNA consists of 2 networks, $G_1$ and $G_2$. The input to the $G_1$ is a batch of few-view sinograms. According to the Fourier slice theorem, low-frequency information is sampled denser than high-frequency information. Therefore, if we perform backprojection directly, reconstructed images will become blurry. Ramp filter is usually used to filter projections for avoiding this blurring effect. In DNA, filtration is performed on the sinogram in the Fourier domain through multiplication with the filter length (filter length, can be shorten, equals twice the length of sinogram). Then, the filtered projections are fed into the first network $G_1$. $G_1$ tries to learn a filtered backprojection algorithm and output an intermediate image. Then, $G_2$ further optimizes the output from $G_1$, generating the final image. 

$G_1$ can be divided into three components: filtration, backprojection, and refinement. In the filtration part, 1-D convolutional layers are used to produce filtered data. Theoretically, the length of the filter is infinitely long for a continuous signal, but it is not practical in reality. Filter length is here set as twice the length of a projection vector (which can be further shortened). Since the filtration is done through a multi-layer CNN, different layers can learn different parts of the filter. Therefore, the 1-D convolutional window is empirically set as $\frac{1}{4}$ the length of the projection vector to reduce the computational burden. The idea of residual connections is used to reserve high-resolution information and to prevent gradient from vanishing. 

Next, the learned sinogram from the filtration part is backprojected by $G_1$. The backprojection part of $G_1$ is inspired by the following intuition: every point in the filtered projection vector only relates to pixel values on the x-ray path through the corresponding object image and any other data points in this vector contribute nothing to the pixels on this x-ray path. There is no doubt that a single fully-connected layer can be implemented to learn the mapping from the sinogram domain to the image domain, but its memory requirement becomes an issue due to extremely large matrix multiplications in this layer. To reduce the memory burden, the reconstruction process is learned in a point-wise manner using a point-wise fully-connected layer. By doing so, DNA can truly learn the backprojection process. The input to the point-wise fully-connected layer is a single point in the filtered projection vector, and the number of neurons is the width of the corresponding image. After this point-wise fully-connected layer, rotation and summation operations are applied to simulate the analytical FBP method. Bilinear interpolation \cite{gribbon_novel_2004} is used for rotating images. Moreover, $C$ is empirically set as 23, allowing the network to learn multiple mappings from the sinogram domain to the image domain. The number of $C$ can be understood as the number of branches. Note that different view-angle uses different parameters. Although the proposed filtration and backprojection parts all together learn a refined FBP method, streak artifacts cannot be eliminated perfectly. An image reconstructed by the backprojection part is fed into the last portion of $G_1$ for refinement. 

Refinement part is a typical U-net \cite{ronneberger_u-net:_2015} with conveying paths and is built with the ResNeXt \cite{xie_aggregated_2016-1} structure. U-net was originally designed for biological images segmentation and had been utilized in various applications. For example, Ref \cite{shan_3-d_2018-1, shan_competitive_2019, chen_low-dose_2017-1} use U-net with conveying paths for CT image denoising, Ref \cite{jin_deep_2017, lee_deep-neural-network-based_2019} for few-view CT problem and Ref \cite{quan_compressed_2018} for compressed sensing MRI, etc. U-net in the DNA contains 4 down-sampling and 4 up-sampling layers, each has a stride of 2 and is followed by a rectified linear unit (ReLU). A $3\times 3$ kernel is used in both convolutional and transpose-convolutional layers. The number of kernels in each layer is 36. To maintain the tensor’s size, zero-padding is used.

$G_2$ uses the same structure as the refinement part in $G_1$. The input to $G_2$ is a concatenation of FBP-result and output from $G_1$. With the use of $G_2$, the network becomes deep.  As a result, the benefits of deep learning can be utilized in this direct mapping for CT image reconstruction.

\subsection{Objective functions}

\begin{figure} [htbp]
   \begin{center}
   \begin{tabular}{c} 
   \includegraphics[width=0.96\textwidth]{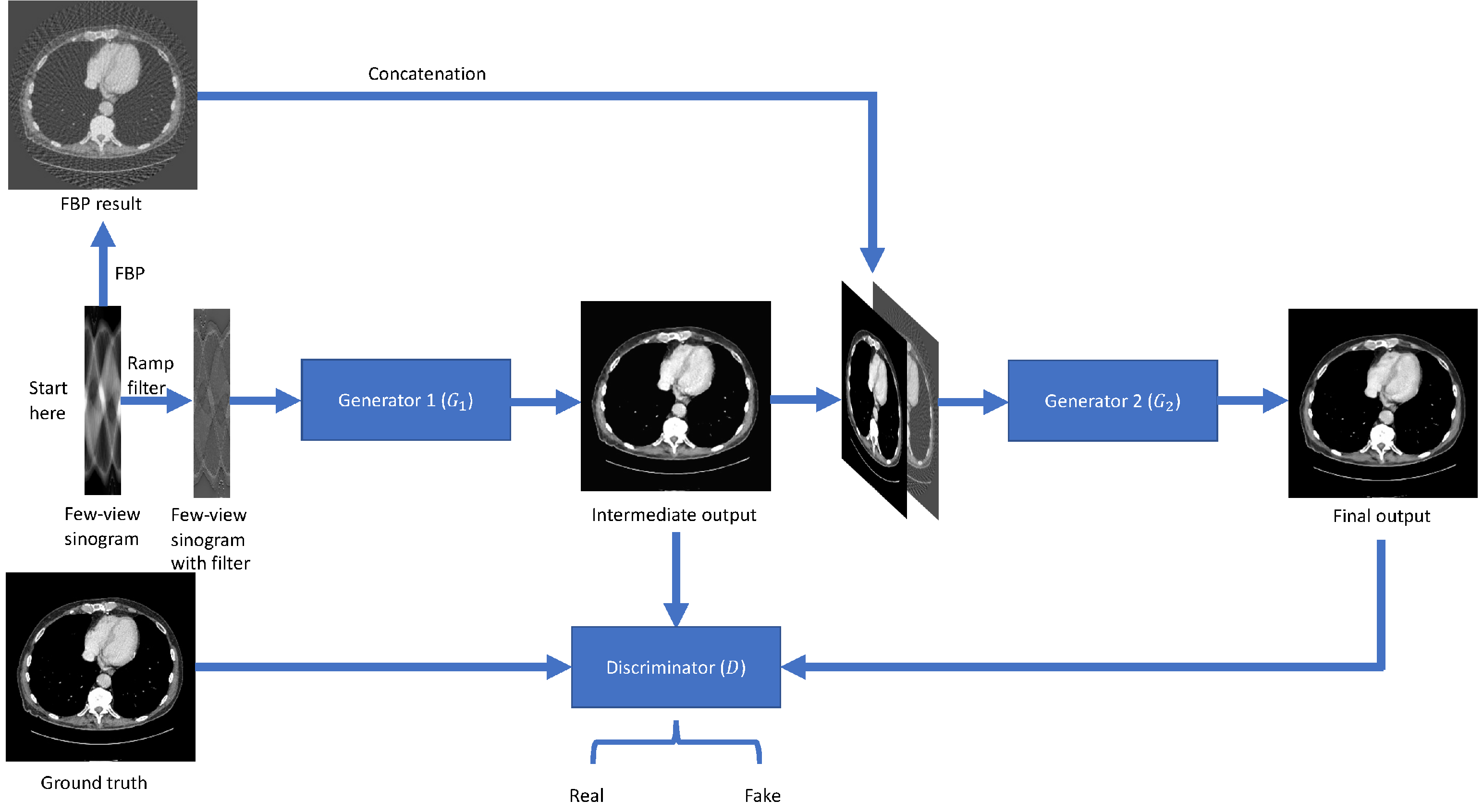}
	\end{tabular}
	\end{center}
   \caption[example] 
   { \label{fig:Network_structure} 
Workflow of the proposed method. Images are example outputs from a 49-view sinogram. The display window is [-160, 240] HU}
   \end{figure}

As shown in Fig.~\ref{fig:Network_structure}, DNA is optimized using the  Generative Adversarial Network (GAN) framework \cite{arjovsky_wasserstein_2017-1}, which is one of the most advanced framework in the field. In this study, the proposed framework contains three components: 2 generator networks $G_1$ and $G_2$ which are introduced in Subsection \ref{sec:2.1}, and a discriminator network $D$. $G_1$ and $G_2$ aim at reconstructing images directly from a batch of few-view sinograms. $D$ receives images from $G_1$, $G_2$ or ground-truth dataset, and intends to distinguish whether an image is real (from the ground-truth dataset) or fake (from either $G_1$ or $G_2$). Both networks are able to optimize themselves in the training process. If an optimized network $D$ can hardly distinguish fake images from real images, we will say that generators $G_1$ and $G_2$ can fool discriminator $D$ which is the goal of GAN. By the design, the network $D$ also helps to improve the texture of the final image and prevent over-smoothed issue from occurring. 

Different from the vanilla generative adversarial network (GAN) \cite{goodfellow_generative_2014}, Wasserstein GAN (WGAN) replaces the logarithm terms in GAN loss function with the Wasserstein distance, improving the training stability during the training process. In the WGAN framework, the 1-Lipschitez function is assumed with weight clipping. However, Ref \cite{gulrajani_improved_2017-1} points out that the weight clipping may be problematic in WGAN and suggests to replace it with a gradient penalty term, which is implemented in our proposed framework. Hence, the objective function of the proposed WGAN framework is expressed as follows:
\begin{equation}
\min_{\theta_{G_1},\theta_{G_2}}\max_{\theta_D}\bigg\{\mathbb{E}_{S_{SV}}\big[D(G_1(S_{SV}))\big]-\mathbb{E}_{I_{FV}}\big[D(I_{FV})\big]-\mathbb{E}_{S_{SV}}\big[D(G_2(I_{SV}))\big]-\mathbb{E}_{I_{FV}}\big[D(I_{FV})\big]+\lambda\mathbb{E}_{\bar{I}}\big[(\|\nabla(\bar{I})\|_2-1)^2\big]\bigg\}
\label{equ:wgan}
\end{equation}
where $S_{SV}$, $I_{SV}=G_1(S_{SV})$, $I_{FV}$ represent a sparse-view sinogram, an image reconstructed by $G_1$ from a sparse-view sinogram  and the ground-truth image reconstructed from the full-vew projection data respectively. $\mathbb{E}_a[b]$ denotes the expectation of $b$ as a function of $a$. $\theta_{G_1}$ ,$\theta_{G_2}$ and $\theta_{D}$ represent the trainable parameters of $G_1$, $G_2$ and $D$ respectively. $\bar{I}$ represents images between fake (from either $G_1$ or $G_2$) and real (from the ground-truth dataset) images. $\nabla(\bar{I})$ denotes the gradient of $D$ with respect to $\bar{I}$. The parameter $\lambda$ balances the Wasserstein distance terms and gradient penalty terms. As suggested in Refs~\cite{arjovsky_wasserstein_2017-1,goodfellow_generative_2014-1, gulrajani_improved_2017-1}, $G_1$, $G_2$ and $D$ are updated iteratively. 

The objective function for optimizing the generator networks involves the mean square error (MSE) \cite{chen_low-dose_2017-1, wolterink_generative_2017}, structural similarity index (SSIM) \cite{zhou_wang_image_2004, you_structurally-sensitive_2018} and adversarial loss \cite{wu_cascaded_2017, yang_low-dose_2018}. MSE is a popular choice for denoising applications \cite{wang_mean_2009}, which effectively suppresses the background noise but could result in over-smoothed images \cite{zhao_loss_2017}. Moreover, MSE is insensitive to image texture since it assumes background noise is white gaussian noise and is independent of local image features. The formula of MSE loss is expressed as follows:
\begin{equation}
\mathcal{L}_2=\frac{1}{N_b\cdot W \cdot H}\sum_{i=1}^{N_b}\|Y_i-X_i\|^2_2
\end{equation}
where $N_b$, $W$ and $H$ denote the number of batches, image width and image height respectively. $Y_i$ and $X_i$ represent ground-truth image and image reconstructed by generator networks (either $G_1$ or $G_2$) respectively.

To compensate for the disadvantages of MSE and acquire visually better images, SSIM is introduced in the objective function. The SSIM formula is expressed as follows:

\begin{equation}
SSIM(Y,X)=\frac{(2\mu_Y\mu_X+C_1)(2\sigma_{YX}+C_2)}{({\mu_Y}^2+{\mu_X}^2+C_1)({\sigma_Y}^2+{\sigma_X}^2+C_2)}
\end{equation}
where $C_1=(K_1\cdot R)^2$ and $C_2=(K_2\cdot R)^2$ are constants used to stabilize the formula if the denominator is small. $R$ stands for the dynamic range of pixel values and $K_1=0.01$, $K_2=0.03$. $\mu_Y$, $\mu_X$, ${\sigma_Y}^2$, ${\sigma_X}^2$ and $\sigma_{XY}$ are the mean of $Y$ and $X$, variance of $Y$ and $X$ and the covariance between $Y$ and $X$ respectively. Then, the structural loss is expressed as follows: 

\begin{equation}
\mathcal{L}_{sl}=1-SSIM(Y,X)
\end{equation}

Furthermore, the adversarial loss aims to assist the generators, producing sharp images that are indistinguishable by the discriminator network. Refer to equation \ref{equ:wgan}, adversarial loss for $G_1$ is expressed as follows:

\begin{equation}
\mathcal{L}^{(1)}_{al}=-\mathbb{E}_{S_{SV}}[D(G_1(S_{SV}))]
\end{equation}
and adversarial loss for $G_2$ is expressed as follows:
\begin{equation}
\mathcal{L}^{(2)}_{al}=-\mathbb{E}_{S_{SV}}\big[D(G_2(I_{SV}))\big]
\end{equation}

As mentioned early, solving the few-view CT problem is similar to solving a set of linear equations when the number of equations is not sufficient to perfectly resolve all the unknowns. The intuition of DNA is trying to estimate those unknown as close as possible by combining the information from the existing equations and the knowledge hidden in the big data. The recovered unknowns should satisfy the equations we have. Therefore, MSE between the original sinogram and the synthesized sinogram from a reconstructed image (either $G_1$ or $G_2$) is also included as part of the objective function, which is expressed as follows:

\begin{equation}
\mathcal{L}_{2}^{sino}=\frac{1}{N_b\cdot V \cdot H}\sum_{i=1}^{N_b}\|Y_{i}^{sino}-X_{i}^{sino}\|^2_2
\end{equation}
where $N_b$, $V,H$ denote the number of batches, number of views and sinogram height respectively. $Y_{i}^{sino}$ represents original sinogram and $X_{i}^{sino}$ represents sinogram from a reconstructed image (either $G_1$ or $G_2$).

Both generator networks are updated at the same time. The overall objective function of 2 generators is expressed as follows:
\begin{equation}
\min_{\theta_{G_1},\theta_{G_2}} [\lambda_Q\cdot (\mathcal{L}^{(1)}_{al} + \mathcal{L}^{(2)}_{al})
+\lambda_P\cdot(\mathcal{L}^{(1)}_{sl}+\mathcal{L}^{(2)}_{sl})
+\lambda_R\cdot (\mathcal{L}_{2}^{sino(1)}+\mathcal{L}_{2}^{sino(2)}) 
+\mathcal{L}^{(2)}_2+\mathcal{L}^{(1)}_2]
\end{equation}
where the superscripts $^{(1)}$ and $^{(2)}$ indicate that the term is for measurements between ground-truth images and results reconstructed by $G_1$ and $G_2$ respectively. $\lambda_Q$, $\lambda_P$ and $\lambda_R$ are hyper-parameters used to balance different loss functions.

\subsection{Discriminator network}
The discriminator network of proposed method takes input from $G_1$, $G_2$, and the ground-truth dataset, trying to distinguish whether the input is real. In DNA, the discriminator network has 6 convolutional layers with 64, 64, 128, 128, 256, 256 filters and followed by 2 fully-connected layers with number of neurons 1,024 and 1 respectively. The leaky ReLU activation function is used after each layer with a slope of 0.2 in the negative part. $3\times 3$ kernel and zero-padding are used for all convolutional layers, with stride equals 1 for odd layers and 2 for even layers.

\section{EXPERIMENTS AND RESULTS}

\subsection{Experimental design}
The clinical patient dataset was generated and authorized by Mayo Clinic for ``\emph{the 2016 NIH-AAPM-Mayo Clinic\ Low Dose CT Grand Challenge}'' \cite{noauthor_low_nodate}. The dataset contains a total of 5,936 abdominal CT images selected by Mayo Clinic with 1 mm slice thickness. Pixel values of patient images were normalized between 0 and 1. In this dataset, 9 patients (5,410 images) are selected for training/validation while 1 patient (526 images) is selected for testing. As mentioned early, DNA was first pre-trained using natural images from ImageNet. During the pre-training segment, a total of 120,000 images were randomly selected from ImageNet, among which 114,000 images were used for training/validation and the remaining 6,000 images were used for testing. Pixel values of ImageNet images were also normalized between 0 and 1. All the pixel values outside a prescribed circle were set to 0. All images were resized into $256\times 256$. The Radon transform was used to simulate few-view projection measurements. 39-view and 49-view sinograms were respectively synthesized from angles equally distributed over a full scan range.

A batch size of 10 was selected for training. All experimental code was implemented in the TensorFlow framework \cite{abadi_tensorflow:_nodate} on an NVIDIA Titan Xp GPU. The Adam optimizer \cite{kingma_adam:_2014} was used to optimize the parameters. We compared DNA with 3 state-of-the-art deep learning methods, including LEARN \cite{chen_learn:_2018}, sinogram-synthesis U-net \cite{lee_deep-neural-network-based_2019}, and iCT-Net \cite{li_learning_2019}. To our best knowledge, the network settings are the same as the default settings described in the original papers. 

For the LEARN network, the number of iterations $\lambda^t$ was set as 50; the number of filters for all three layers was set to 48, 48, and 1 respectively; the convolutional kernel was set as $5\times 5$; the initial input to the network was the FBP result; the same preprocessed Mayo dataset described above was used to train the LEARN network. Please note that the amount of data we used to train the LEARN network was much larger than that in the original LEARN paper. 

For the sinogram-synthesis U-net, 720-view sinograms were simulated using the Radon transform. Then, the simulated sinograms were cropped into $50\times 50$ patches with a stride 10 for training. The FBP method was applied to the reconstructed sinograms for generating final images. 

The training process of iCT-Net is divided into two stages. In the first stage, the first 9 layers were pre-trained with projection data. In the second stage, the pre-trained iCT-Net performed end to end training using the patient data. In the original iCT-Net paper, iCT-Net used a total of 58 CT examinations acquired under the same condition for stage 2 training. However, since we do not have their dataset, limited Mayo images might have caused overfitting in stage 2 training when we made efforts to replicate their results. Therefore, for a fair comparison, testing images were included in the training stages. Please note that iCT-Net was handled by 2 NVIDIA Titan Xp GPUs.

\subsection{Visual and quantitative assessment}
\begin{figure} [htbp]
   \begin{center}
   \begin{tabular}{c} 
   \includegraphics[width=0.89\textwidth, trim={1cm, 1.5cm, 8.5cm, 1cm}]{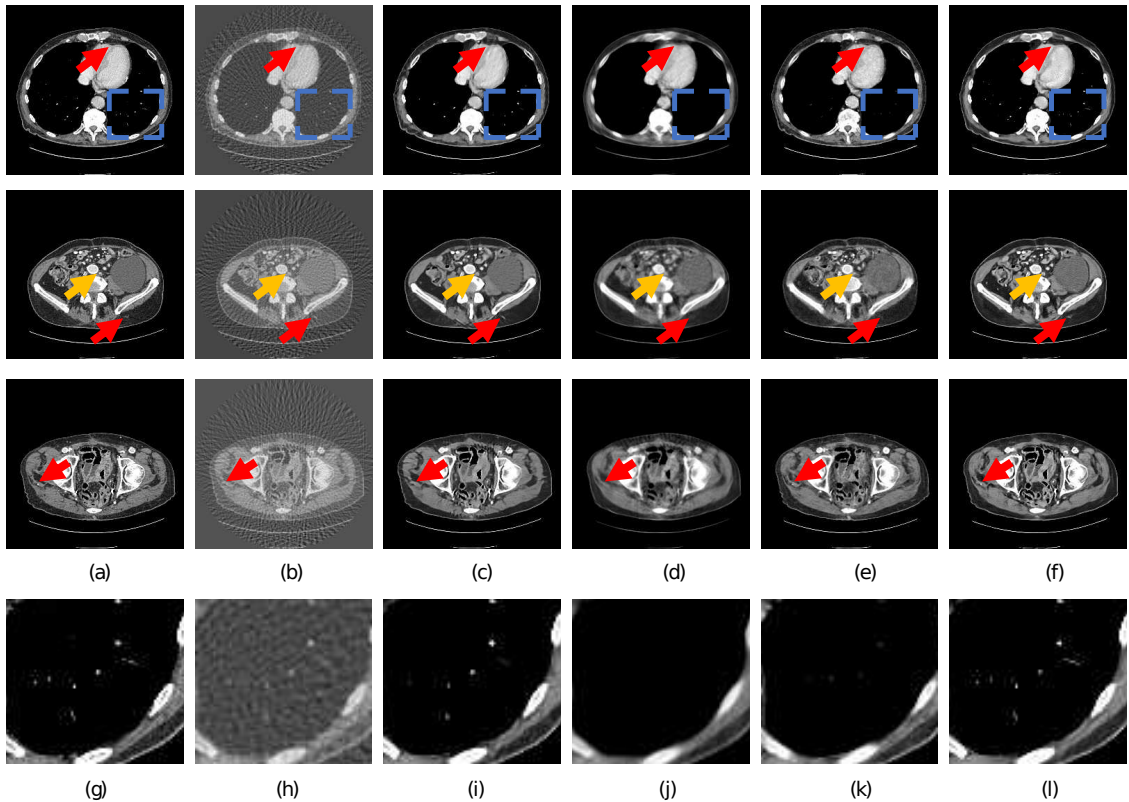}
	\end{tabular}
	\end{center}
   \caption[example] 
   { \label{fig:Mayo_testing} 
Representative images reconstructed using different methods. The display window is [-160, 240] HU for patient images. (a) The ground-truth, (b) FBP, (c) LEARN, (d) sinogram-synthesis U-net, (e) iCT-Net, (f) DNA. (g)-(l) The zoomed region of the first row marked by the blue boxes. More tiny details were recovered using our proposed method. The red arrows point to some small details that are better reconstructed by DNA.}
   \end{figure}

To visualize the performance of different methods, a few representative slices were selected from the testing dataset. Figure \ref{fig:Mayo_testing} shows results reconstructed using different methods from 49-view real patient sinograms. Table \ref{tab:Num_parameters} shows the number of parameters used for these methods.

\begin{table}[htbp]
\caption{The number of parameters used in different methods.} 
\label{tab:Num_parameters}
\begin{center}       
\begin{tabular}{|c|c|c|c|c|c|} 
\hline
\rule[-1ex]{0pt}{3.5ex}   & LEARN & Sino-synthesis U-net & iCT-Net & DNA (49-views) & DNA (39-views)  \\
\hline
\rule[-1ex]{0pt}{3.5ex} No. of parameters & 3,004,900 & 47,118,017 & 16,933,929 & 1,962,101 & 1,844,341   \\
\hline
\end{tabular}
\end{center}
\end{table}

Three metrics, PSNR, SSIM, and root-mean-square-error (RMSE), are selected for quantitative assessment. Table \ref{tab:quan_different_methods} shows quantitative measurements for different methods, acquired by averaging the metrics over the testing dataset, for both 39-view and 49-view results. 

\begin{table}[htbp]
\caption{Quantitative measurements for different methods ($MEAN\pm STD$). For each metric, the best results are marked in red. We did not test the iCT-Net for the 39-view case. Measurements are acquired by averaging the values in the testing dataset.} 
\label{tab:quan_different_methods}
\begin{center}       
\begin{tabular}{|c|c|c|c|c|} 
\hline
 & LEARN & Sino-syn U-net & iCT-Net & DNA \\
\hline
\rule[-1ex]{0pt}{3.5ex}  SSIM (49-views) & $0.900\pm 0.026$ & $0.814 \pm 0.029$ & $0.784 \pm 0.020$ & \textcolor{red}{$0.913 \pm 0.023$}   \\
\hline
\rule[-1ex]{0pt}{3.5ex}  PSNR (49-views) & $28.966 \pm 1.262$ & $24.858 \pm 0.777$ & $27.062 \pm 0.904$ & \textcolor{red}{$29.174\pm 1.234$}   \\
\hline
\rule[-1ex]{0pt}{3.5ex}  RMSE (49-views) & $0.036 \pm 0.006$ & $0.057 \pm 0.005$ & $0.045 \pm 0.005$ & \textcolor{red}{$0.035\pm 0.005$}   \\
\hline
\rule[-1ex]{0pt}{3.5ex}  SSIM (39-views) & $0.882 \pm 0.029$ & $0.781 \pm 0.029$ & N/A & \textcolor{red}{$0.897 \pm 0.026$}   \\
\hline
\rule[-1ex]{0pt}{3.5ex}  PSNR (39-views) & $27.994 \pm 1.211$ & $24.040 \pm 0.753$ & N/A & \textcolor{red}{$28.294 \pm 1.260$}   \\
\hline
\rule[-1ex]{0pt}{3.5ex}  RMSE (39-views) & $0.040 \pm 0.006$ & $0.063 \pm 0.006$ & N/A & \textcolor{red}{$0.038 \pm 0.006$}   \\
\hline

\end{tabular}
\end{center}
\end{table}

\subsection{Examination of intermediate results}
\begin{figure} [htbp]
   \begin{center}
   \begin{tabular}{c} 
   \includegraphics[width=0.96\textwidth]{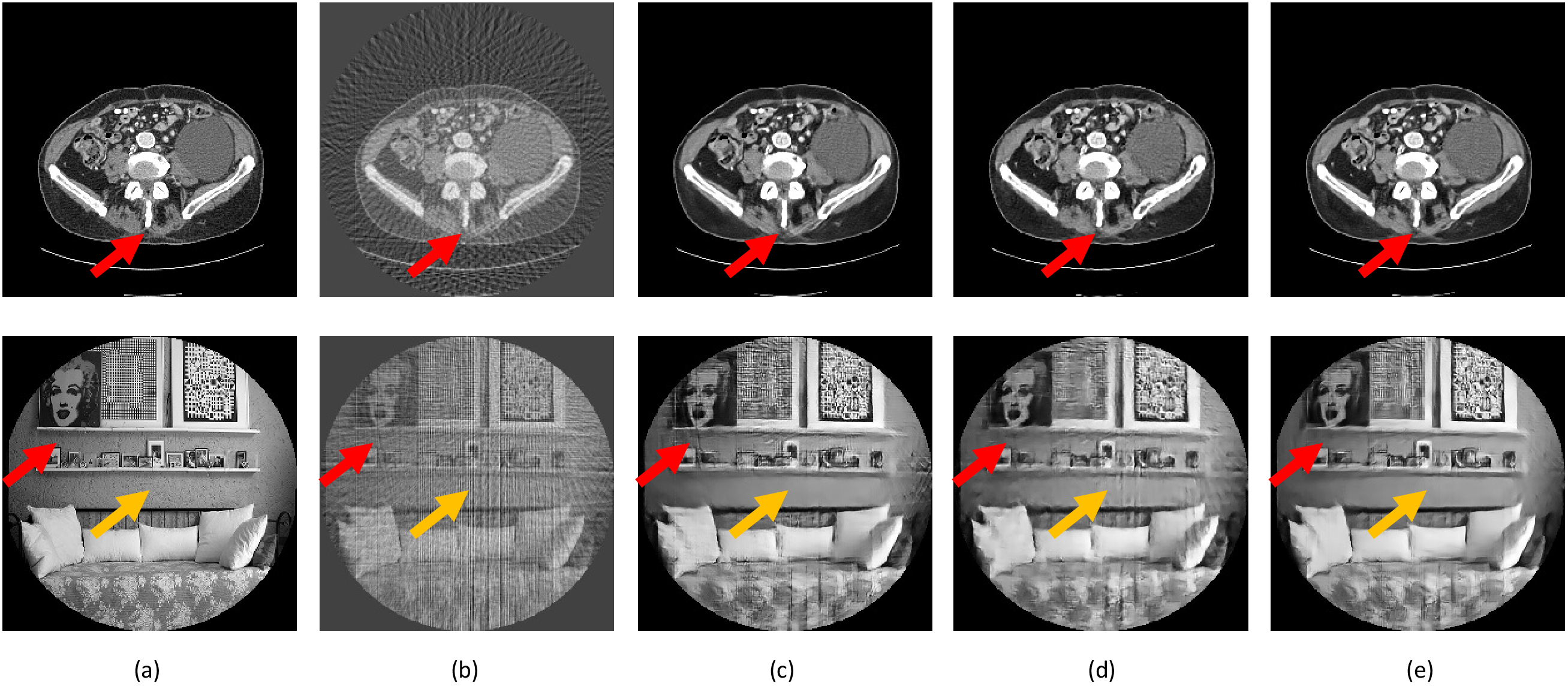}
	\end{tabular}
	\end{center}
   \caption[example] 
   { \label{fig:intermidiate_test} 
Representative outputs using different methods. (a) The ground-truth, (b) FBP, (c) $G_2$ trained using only FBP results as the input, (d) $G_1$, and (e) DNA. The display window is [-160, 240] HU for patient images.}
   \end{figure}

To demonstrate the effectiveness of two generators in DNA, another $G_2$ network was trained using solely the FBP results as the input. Figure \ref{fig:intermidiate_test} shows typical results reconstructed from 49-view sinograms using various methods.

As shown in the first row in Figure \ref{fig:intermidiate_test}, streak artifacts cannot be perfectly eliminated by $G_2$ when the network is trained using only the FBP results. $G_1$, on the other hand, eliminates these artifacts through learning from projection data (first row, red arrows). Moreover, as mentioned early, by the design $G_1$ intends to assist $G_2$ for producing better results. This effect can be observed in the second row of Figure \ref{fig:intermidiate_test}. $G_1$ removes artifacts that cannot be removed using $G_2$ (second row, red arrow), but it introduces new artifacts (second row, orange arrow). These artifacts can then be removed by $G_2$. In summary, $G_1$ helps remove artifacts that could not be removed by processing FBP results. Even though it brings up new artifacts, the newly introduced artifacts can be removed by the second generator network. Put differently, the proposed method, DNA, combines the best of two worlds. Quantitative measurements on the outputs reconstructed by various components in DNA are listed in Table \ref{tab:quan_intermidiate_DNA}. 

\begin{table}[ht]
\caption{Quantitative measurements for different components in DNA ($MEAN\pm STD$). The best results are marked in red. Measurements are acquired by averaging the values in the testing dataset.} 
\label{tab:quan_intermidiate_DNA}
\begin{center}       
\begin{tabular}{|c|c|c|c|} 
\hline
\rule[-1ex]{0pt}{3.5ex}   & $G_1$ & $G_2$(trained using only FBP results) & DNA \\
\hline
\rule[-1ex]{0pt}{3.5ex}  SSIM (49-views) & $0.899\pm 0.025$ & $0.908\pm 0.023$ & \textcolor{red}{$0.913\pm 0.023$}   \\
\hline
\rule[-1ex]{0pt}{3.5ex}  PSNR (49-views) & $28.015\pm 1.111$ & $28.789\pm 1.167$ & \textcolor{red}{$29.174\pm 1.234$}   \\
\hline
\rule[-1ex]{0pt}{3.5ex}  RMSE (49-views) & $0.040\pm 0.005$ & $0.037\pm 0.005$ & \textcolor{red}{$0.035\pm 0.005$}   \\
\hline

\end{tabular}
\end{center}
\end{table}

\subsection{Generalizability analysis}
\begin{figure} [ht]
   \begin{center}
   \begin{tabular}{c} 
   \includegraphics[width=0.96\textwidth]{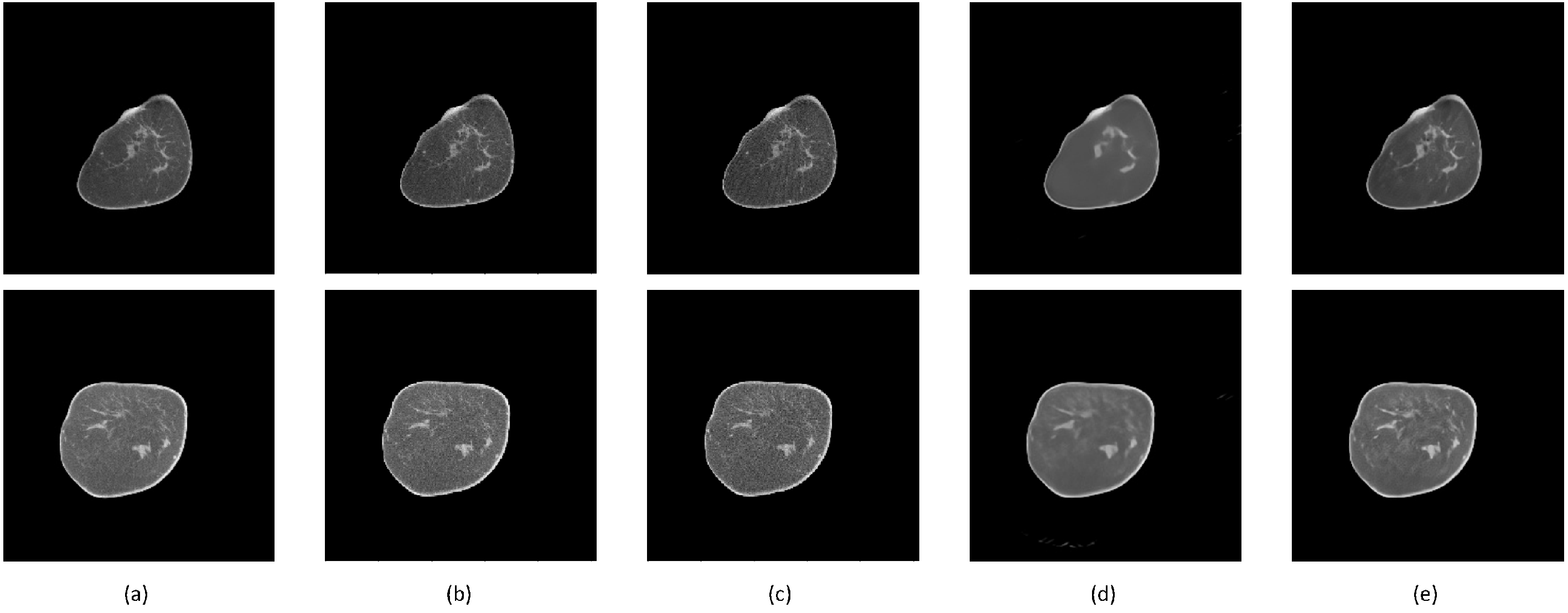}
	\end{tabular}
	\end{center}
   \caption[example] 
   { \label{fig:Koning_test} 
Representative outputs using different methods on Koning breast CT datasets. (a) Koning scanner (300-view), (b) Koning scanner (150-view), and (c) Koning scanner (100-view) (d) LEARN (49-view) and (e) DNA (49-view). The display window is [-300, 300] HU.}
   \end{figure}

To demonstrate that the proposed method truly learns the backprojection process and can be generalized to other datasets, DNA and LEARN (second best method) were tested directly on female breast CT datasets acquired on a breast CT scanner (Koning corporation). 4,968 CT images, scanned at 49 peak kilovoltage (kVp), were acquired from 12 patients. There is a total of 3 sets of images per patient, reconstructing from 300, 150, 100 projections respectively. Koning images are reconstructed using commercial FBP with additional post-processing. Figure \ref{fig:Koning_test} shows representative images reconstructed using different methods. Table \ref{tab:quan_koning} gives the corresponding quantitative measurements. Completely dark images in the dataset were excluded, resulting in a total of 4,635 CT images. 

DNA demonstrates outstanding performance in terms of generalizability, as shown in Figure \ref{fig:Koning_test}. Specifically, images reconstructed using LEARN appear over-smoothed and lose some details. On the other hand, DNA not only reserves more details than LEARN does, but also suppresses streak artifacts effectively (relative to 150-view and 100-view results). Moreover, images reconstructed by DNA from 49-view sinograms are better than 100-view images in terms of SSIM and RMSE.

\begin{table}[ht]
\caption{Quantitative measurements for Koning breast images reconstructed using different methods ($MEAN\pm STD$). Measurements were calculated with respect to 300-view results and acquired by averaging the values in the testing dataset. The best and second-best results are marked in red and blue respectively.} 
\label{tab:quan_koning}
\begin{center}       
\begin{tabular}{|c|c|c|c|c|} 
\hline
\rule[-1ex]{0pt}{3.5ex}   & Koning commercial FBP & Koning commercial FBP & LEARN (49-view) & DNA (49-view)\\
 & (150-view) & (100-view) & &\\
\hline
\rule[-1ex]{0pt}{3.5ex}  SSIM &  \textcolor{red}{$0.972\pm 0.038$} & $0.953\pm 0.062$ & $0.922\pm 0.076$ & \textcolor{blue}{$0.957\pm 0.057$}   \\
\hline
\rule[-1ex]{0pt}{3.5ex}  PSNR &  \textcolor{red}{$42.455\pm 10.141$} &  \textcolor{blue}{$38.880\pm 10.022$} & $34.588\pm 3.658$ & $37.415\pm 4.259$  \\
\hline
\rule[-1ex]{0pt}{3.5ex}  RMSE &  \textcolor{red}{$0.012\pm 0.010$} & $0.018\pm 0.015$ & $0.021\pm 0.011$ & \textcolor{blue}{$0.015\pm 0.008$}  \\
\hline

\end{tabular}
\end{center}
\end{table}

\begin{figure} [ht]
   \begin{center}
   \begin{tabular}{c} 
   \includegraphics[width=0.96\textwidth]{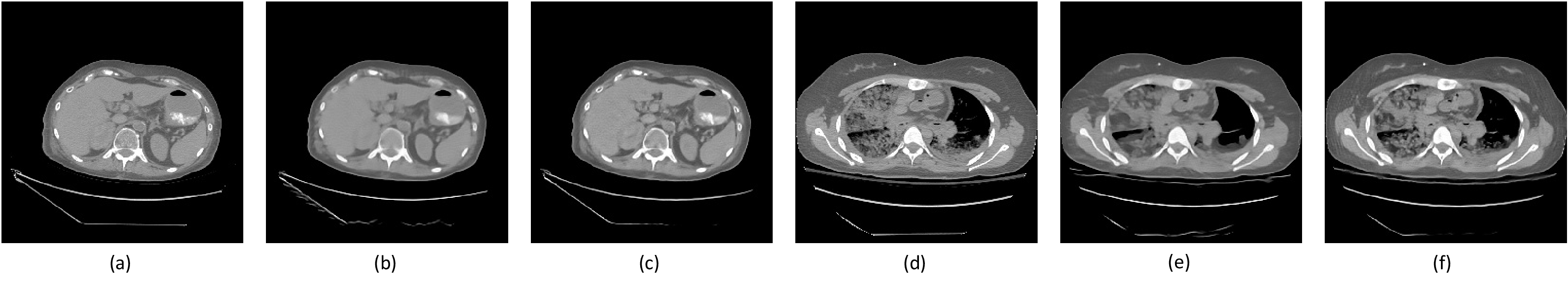}
	\end{tabular}
	\end{center}
   \caption[example] 
   { \label{fig:MGH_test} 
Representative outputs using different methods for the MGH dataset from 49-view sinograms. (a) and (d) Ground-truth, (b) and (e) LEARN, (c) and (f) DNA. The display window is [-300, 300] HU.}
   \end{figure}

Also, we validated DNA and LEARN on the Massachusetts General Hospital (MGH) dataset \cite{yang_big_2015}. MGH dataset contains 40 cadaver scans acquired with representative protocols. Each cadaver was scanned on a GE Discovery HD 750 scanner at 4 different dose levels. Only 10\textit{NI} (Noise Index) images were used for testing. \textit{NI} refers to the standard deviation of CT numbers within a region of interest in a water phantom of a specific size \cite{mccollough_ct_2006}. Typical images are shown in Figure \ref{fig:MGH_test}. The corresponding quantitative measurements are shown in Table \ref{tab:quan_MGH}.

\begin{table}[h]
\caption{Quantitative measurements for the MGH dataset ($MEAN\pm STD$). Measurements were acquired by averaging the values in the testing dataset. The best results are marked in red.} 
\label{tab:quan_MGH}
\begin{center}       
\begin{tabular}{|c|c|c|} 
\hline
\rule[-1ex]{0pt}{3.5ex}   & LEARN (49-view) & DNA (49-view)\\
\hline
\rule[-1ex]{0pt}{3.5ex}  SSIM &  $0.849\pm 0.039$ &  \textcolor{red}{$0.874\pm 0.039$}   \\
\hline
\rule[-1ex]{0pt}{3.5ex}  PSNR &  $26.193\pm 1.131$ & \textcolor{red}{$27.687\pm 1.566$}  \\
\hline
\rule[-1ex]{0pt}{3.5ex}  RMSE &  $0.049\pm 0.007$ & \textcolor{red}{$0.042\pm 0.008$}  \\
\hline

\end{tabular}
\end{center}
\end{table}

\section{CONCLUSION}
Although the field of deep imaging is still at its early stage, remarkable results have been achieved over the past several years. We envision that deep learning will play an important role in the field of tomographic imaging \cite{wang_machine_2017}. In this direction, we have developed this novel DNA network for reconstructing CT images directly from sinograms. In this paper, even though the proposed method has only been tested for the few-view CT problem, we believe that it can be applied/adapted to solve various other CT problems, including image de-noising, limited-angle CT, and so on. This paper is not the first work for reconstructing images directly from raw data, but previously proposed methods require a significantly greater amount of GPU memory for training. It is underlined that our proposed method solves the memory issue by learning the reconstruction process with the point-wise fully-connected layer and other proper network ingredients. Also, by passing only a single point into the fully-connected layer, the proposed method can truly learn the backprojection process. In our study, the DNA network demonstrates superior performance and generalizability. In the future works, we will validate the proposed method on images up to dimension $512\times 512$ or even $1024\times 1024$.

% References
\bibliography{main} % bibliography data in report.bib
\bibliographystyle{spiebib} % makes bibtex use spiebib.bst

\end{document}